%% file: main.tex
\documentclass[12pt]{article}

\usepackage{amsmath}
\usepackage{amssymb}
\usepackage{authblk}
\usepackage{booktabs}
\usepackage{dsfont}
\usepackage{float}
\usepackage[T1]{fontenc}
\usepackage[margin=1in]{geometry}
\usepackage{graphicx}
\usepackage{hyperref}
\usepackage{microtype}
\usepackage{natbib}
\usepackage{parskip}
\usepackage{xcolor}

\hypersetup{
  colorlinks=true,
  citecolor=blue!60!black,
  linkcolor=blue!60!black,
  urlcolor=blue!60!black
}
\makeatletter
\renewcommand{\@seccntformat}[1]{%
  \csname the#1\endcsname\hspace{0.75em}%
}
\makeatother

\input{results.tex}

\title{Dummy RAPM: Representing Low-Minute Players in\\
Regularized Adjusted Plus-Minus}
\author{Kenny Watts}
\author{Jonathan Pipping-Gam\'on}
\author{Abraham J. Wyner}
\affil{\small The Wharton School, University of Pennsylvania}
\date{July 2026}

\begin{document}

\maketitle

\begin{abstract}
Regularized Adjusted Plus-Minus (RAPM) uses stint-level lineup indicators to
estimate player contributions to scoring margin. When low-minute player
columns are removed, their stints remain in the data, but the design matrix no
longer represents the complete lineup. Dummy RAPM restores this information
using five indicators for the number of excluded players on each lineup side.
Across \NumberSeasons{} NBA seasons, chronological validation selects a
\SelectedThreshold-minute-per-appearance threshold and a dummy-to-player
penalty ratio of \SelectedPenaltyRatio. On held-out March--April games, Dummy
RAPM reduces mean season game-margin RMSE from \MeanNoDummyRMSE{} to
\MeanDummyRMSE{} and achieves lower RMSE in \RMSEWins{} of \NumberSeasons{}
seasons. The average reduction is \MeanRMSEGain{} points, or
\RelativeRMSEGain{}. Although the improvement in game-level predictive
accuracy is small, it is consistent: RAPM performs better when it records how
many excluded players are on each side.
\end{abstract}

\section{Introduction}

Basic plus-minus measures a team's score differential while a player is on the
court but does not adjust for teammates or opponents. Adjusted Plus-Minus
(APM), associated with the WINVAL work of Jeff Sagarin and Wayne Winston and
described publicly by \citet{Rosenbaum2004}, addresses this limitation by
regressing stint-level score differentials on player indicators.

APM estimates are unstable when players repeatedly share the floor with the
same teammates and opponents or appear in few stints. Ridge regression
stabilizes correlated designs by shrinking coefficients
\citep{HoerlKennard1970}. \citet{Sill2010} applied this approach to APM and
emphasized evaluation on future games, yielding Regularized Adjusted
Plus-Minus (RAPM). Box Plus/Minus represents a separate, complementary line of
work based on box-score and role information \citep{Myers2020}. Bayesian
partial-pooling approaches provide a related response to weak identification.
\citet{FearnheadTaylor2011} estimated offensive and defensive abilities while
borrowing information across seasons, and \citet{DeshpandeJensen2016} modeled
changes in win probability using Bayesian linear regression.

Some RAPM implementations remove low-minute players before fitting. Removing
these columns leaves their stints in the data but makes the lineup
representation incomplete. The model may then attribute omitted contributions
to retained players whose indicators are correlated with the missing lineup
information.

Dummy RAPM addresses this omission without estimating a separate coefficient
for every low-minute player. It records how many excluded players appear on
the home and away sides, then pools those count coefficients with ridge
regularization. The question is simple: does restoring this small piece of
lineup information improve prediction?

\section{Model}

\subsection{Stint response and player design}

A stint is an interval over which neither lineup changes. Let $\Delta_i$ be the
change in home-minus-away score margin during stint $i$, and define exposure
\[
q_i = \frac{P_{H,i}+P_{A,i}}{2},
\]
where $P_{H,i}$ and $P_{A,i}$ count home and away possession-ending events.
The response is
\[
y_i = 100\frac{\Delta_i}{q_i},
\]
the home-team margin per 100 team possessions. Observation $i$ receives weight
$q_i$, so a longer stint contributes more to estimation. Scaling before
estimation makes coefficients directly interpretable in points per 100 team
possessions.

For retained player $j$, $x_{ij}=1$ when the player is in the home lineup,
$x_{ij}=-1$ when in the away lineup, and $x_{ij}=0$ otherwise. Filtered RAPM
estimates
\[
\left(\hat{\alpha},\hat{\boldsymbol{\beta}}\right)
=
\arg\min_{\alpha,\boldsymbol{\beta}}
\left\{
\sum_{i=1}^{n}q_i
\left(y_i-\alpha-\mathbf{x}_i^\mathsf{T}\boldsymbol{\beta}\right)^2
+\lambda_P\|\boldsymbol{\beta}\|_2^2
\right\}.
\]
The intercept is not penalized, and predictors are not standardized. Ridge
solutions are computed over a decreasing sequence of $\lambda_P$ values using
cyclic coordinate descent \citep{FriedmanHastieTibshirani2010}.

\subsection{Lineup-side count indicators}

For $k=1,\ldots,5$, define
\[
d_{H,i}^{(k)}
=
\mathds{1}\{\text{the home lineup in stint }i
\text{ contains exactly }k\text{ excluded players}\},
\]
\[
d_{A,i}^{(k)}
=
\mathds{1}\{\text{the away lineup in stint }i
\text{ contains exactly }k\text{ excluded players}\}.
\]
Lineups with no excluded players form the reference category. The augmented
model therefore contains ten dummy regressors, five for each lineup side.
Separate home and away coefficient blocks allow the association for a given
count to differ according to which side contains the excluded players.

Let $\boldsymbol{\gamma}_H$ and $\boldsymbol{\gamma}_A$ denote the two
five-element dummy blocks. Dummy RAPM estimates
\[
\begin{split}
\arg\min_{\alpha,\boldsymbol{\beta},
\boldsymbol{\gamma}_H,\boldsymbol{\gamma}_A}
\quad&
\sum_{i=1}^{n}q_i
\left(
y_i-\alpha-\mathbf{x}_i^\mathsf{T}\boldsymbol{\beta}
-\mathbf{d}_{H,i}^\mathsf{T}\boldsymbol{\gamma}_H
-\mathbf{d}_{A,i}^\mathsf{T}\boldsymbol{\gamma}_A
\right)^2\\
&+\lambda_P\|\boldsymbol{\beta}\|_2^2
+\lambda_D\left(
\|\boldsymbol{\gamma}_H\|_2^2+
\|\boldsymbol{\gamma}_A\|_2^2
\right).
\end{split}
\]
The relative penalty $\rho=\lambda_D/\lambda_P$ is tuned from the data. A
common penalty for the dummy coefficients allows them to be pooled more or
less strongly than the player coefficients without requiring ten separate
hyperparameters.

\section{Data Construction}

We use publicly accessible ESPN play-by-play and player box-score records
distributed through the sportsdataverse project \citep{Gilani2026}. Starter
flags identify each opening lineup, and recorded substitutions divide the game
into stints with unchanged lineups. We retain a game only when both teams have
five unique recorded starters and every modeled stint contains five unique
home players and five unique away players. This removes
\StarterExcludedGames{} games, including
\StarterExcludedTwentyNineteenGames{} in 2019.

Stint scoring is reconstructed from changes in the reported home-minus-away
scoreboard margin. This captures technical and other free throws even when the
corresponding event is not marked as ending a possession. A scoring segment
with no possession exposure is merged with the next positive-exposure stint,
or with the preceding stint at the end of a game. We verify that the resulting
stint margins sum to the official final margin. Within each game, we retain the
contiguous sequence of periods beginning with period 1, preserving valid
multi-overtime games while discarding disconnected records. Garbage-time and
overtime possessions remain in the sample so that stint predictions aggregate
to the full-game margin.

Player playing time is summarized by mean minutes per appearance within the
relevant training window. This reflects a player's role when active rather
than the number of games he happened to play. As in \citet{Rosenbaum2004},
low-minute players are pooled rather than estimated individually, but we
select the cutoff by chronological validation instead of fixing it at 250
total minutes across two seasons.

Table~\ref{tab:sample} reports the season-level sample. The analysis contains
\TotalModeledGames{} games and \TotalModeledStints{} positive-exposure stints.
The ``Out'' column counts regular-season games in May--September, outside the
analysis windows; these games may also appear among the starter exclusions.

\begin{table}[htbp]
\centering
\caption{Season-level analysis sample and chronological windows}
\label{tab:sample}
\input{sample_summary_table.tex}
\end{table}

\section{Selection and Evaluation}

The sample contains \NumberSeasons{} seasons: 2007--2011, 2013--2019, and
2022--2025. Seasons are labeled by the calendar year in which they end. The
series begins in 2007 because starter metadata are incomplete in earlier
seasons. The lockout-shortened 2011--12 season, disrupted 2019--20 season, and
COVID-shortened 2020--21 season are excluded because their schedules do not
support the common October--April chronological split.

Within each season, October--December games form the inner training period,
January--February games form the inner validation period, and March--April
games form the outer test period. Outer training is exactly the union of the
October--December and January--February windows. May--September games are in
none of the four analysis sets. The windows are disjoint and strictly
chronological.

The initial inner search considers mean-minutes-per-appearance thresholds
\[
\{0,5,10,15,20,25,30\}
\]
and dummy-to-player penalty ratios
\[
\{0,0.1,0.25,0.5,1,2\}.
\]
Threshold zero retains every training-observed player and serves as the
unfiltered RAPM benchmark. The cutoff determines which player columns are
pooled and is chosen solely by predictive performance.

In the first stage, we evaluate every threshold-ratio pair in the initial grid.
For each pair, we average the inner-validation RMSEs of Dummy RAPM and filtered
RAPM, then average that score across seasons. The minimizing pair selects a
threshold of \SelectedThreshold{} minutes per appearance. Holding that
threshold fixed for both models, the second stage evaluates the dummy-to-player
penalty ratio over
\[
\rho\in\{1.0,1.1,\ldots,2.5\}.
\]
If the minimum occurs at a grid boundary, we extend the grid in increments of
0.1 until the selected value is interior. The checked grid runs from
\CheckedPenaltyMinimum{} to \CheckedPenaltyMaximum{} and selects
$\rho=\SelectedPenaltyRatio$.
Both stages minimize this model-averaged, season-averaged inner-validation
RMSE. Exact ties prefer the lower minute threshold and then the larger dummy
penalty. March--April results are not used for selection.

After selection, both models are refit using all games through February and
evaluated once on March--April games. Player eligibility is recomputed using
only the corresponding training period. For each season and training window,
we construct one deterministic five-fold partition at the game level and use
it for every threshold, model, and penalty ratio. Within each fit, $\lambda_P$
minimizes cross-validated game-margin RMSE.

For game $g$, stint predictions are converted back to points and summed:
\[
\widehat{\Delta}_g
=
\sum_{i\in g}\widehat{y}_i q_i/100.
\]
For season $s$, game-margin RMSE is
\[
\operatorname{RMSE}_s
=
\sqrt{
\frac{1}{|\mathcal{T}_s|}
\sum_{g\in\mathcal{T}_s}
(\Delta_g-\widehat{\Delta}_g)^2
}.
\]
Held-out predictive fit is summarized by
\[
R^2_{\mathrm{oos},s}
=
1-\frac{\sum_{g\in\mathcal{T}_s}
(\Delta_g-\widehat{\Delta}_g)^2}
{\sum_{g\in\mathcal{T}_s}
(\Delta_g-\overline{\Delta}_{\mathcal{T}_s})^2},
\]
where $\mathcal{T}_s$ is the March--April test set for season $s$ and
$\overline{\Delta}_{\mathcal{T}_s}$ is its mean observed game margin. The
denominator is the test-set total sum of squares. Mean season RMSE and mean
season $R^2_{\mathrm{oos}}$ are arithmetic averages of these season-specific
quantities, giving each season equal weight rather than pooling all test games.

The paired $t$-test on season-level RMSE differences is the primary
inferential analysis. The Wilcoxon signed-rank and sign tests are rank-based
and direction-only robustness checks; the unfiltered-benchmark and
2019-exclusion comparisons are secondary sensitivity analyses. These nominal
tests treat seasons as independent observational units. Because adjacent
seasons share players, coaches, and league conditions, their $p$-values are
best interpreted as approximate across-season summaries.

Because predictions condition on the realized lineups in the held-out games,
the target is lineup-conditioned final margin rather than a pregame forecast.

\begin{figure}[htbp]
\centering
\includegraphics[width=0.78\linewidth]{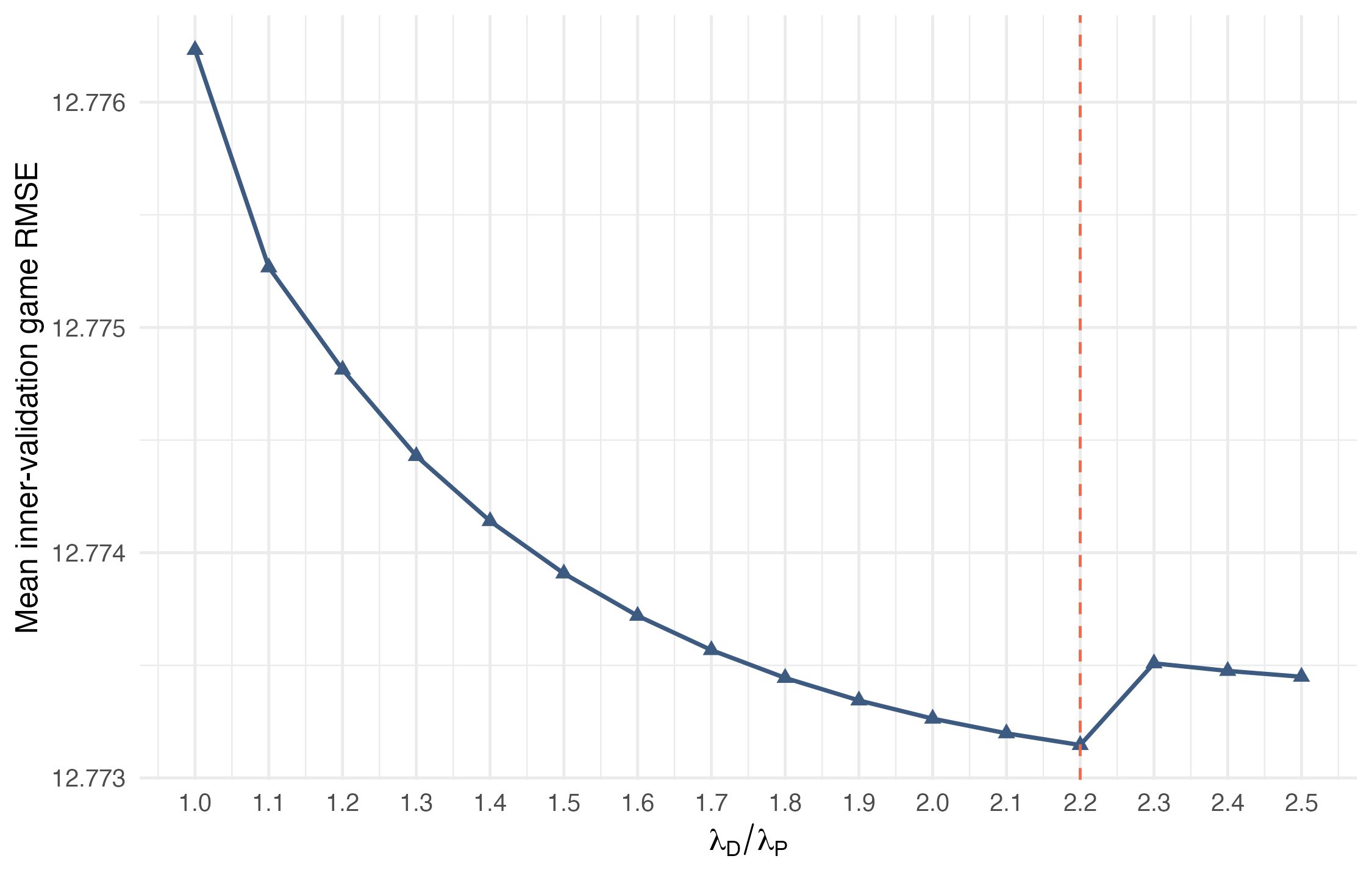}
\caption{Mean inner-validation game-margin RMSE over the dummy-to-player
penalty grid. The dashed line marks the selected ratio,
\SelectedPenaltyRatio.}
\label{fig:penalty}
\end{figure}

\section{Results}

\subsection{Held-out prediction}

\begin{table}[H]
\centering
\caption{Outer-test predictive performance across \NumberSeasons{} seasons}
\label{tab:prediction}
\begin{tabular}{lccc}
\toprule
Model & Mean season RMSE & Mean season $R^2_{\mathrm{oos}}$ &
Lower-RMSE seasons \\
\midrule
Dummy RAPM & \MeanDummyRMSE & \MeanDummyRSquared & \RMSEWins \\
Filtered RAPM & \MeanNoDummyRMSE & \MeanNoDummyRSquared &
\NoDummyRMSEWins \\
\bottomrule
\end{tabular}
\end{table}

Dummy RAPM reduces mean season game-margin RMSE from \MeanNoDummyRMSE{} to
\MeanDummyRMSE{}. This reduction of \MeanRMSEGain{} points, equivalent to
\RelativeRMSEGain{} of the filtered RAPM RMSE, occurs in \RMSEWins{} of
\NumberSeasons{} seasons.
Mean season $R^2_{\mathrm{oos}}$ rises from \MeanNoDummyRSquared{} to
\MeanDummyRSquared{}. Thus, the count indicators produce a small but consistent
improvement in predictive accuracy across seasons.

The across-season 95\% $t$-based confidence interval for the reduction in RMSE is
\RMSEGainCILow{} to \RMSEGainCIHigh{} points. The paired $t$-test gives
$t(\PairedTDegreesFreedom)=\PairedTStatistic$ and $p=\PairedTPValue$.
As robustness checks, the Wilcoxon signed-rank test gives
$W=\WilcoxonStatistic$ and
$p=\WilcoxonPValue$. The sign test records lower RMSE in
\SignTestSuccesses{} of \NumberSeasons{} seasons ($p=\SignTestPValue$).

\begin{figure}[htbp]
\centering
\includegraphics[width=0.82\linewidth]{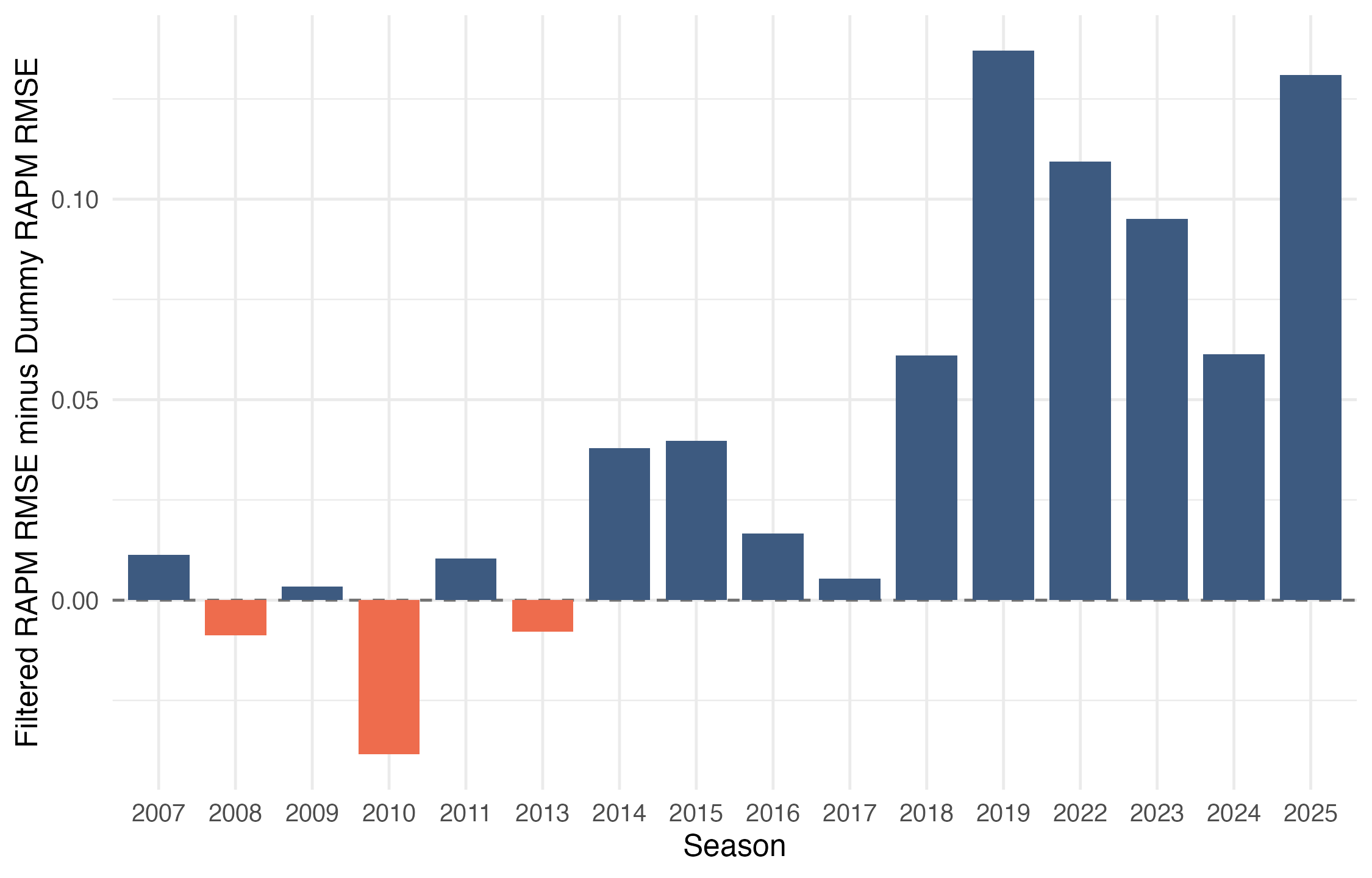}
\caption{Outer-test game-margin RMSE difference by season. Positive values
indicate lower RMSE for Dummy RAPM.}
\label{fig:rmsegain}
\end{figure}

As a secondary sensitivity analysis, Dummy RAPM has lower RMSE in
\PreTwentyTwentyWins{} of
\PreTwentyTwentySeasons{} seasons through 2019 and in
\PostCovidWins{} of \PostCovidSeasons{} seasons from 2022 onward.
Excluding 2019, the season with the most starter-data exclusions, leaves
\NumberSeasonsExcludingTwentyNineteen{} seasons with an average RMSE reduction of
\ExcludingTwentyNineteenRMSEGain{} points
($p=\ExcludingTwentyNineteenPValue$).

\subsection{Filtering and representation}

The unfiltered benchmark provides a second sensitivity analysis. Filtering
low-minute players without adding count indicators does not improve RAPM.
Relative to the unfiltered benchmark, the selected filtered model has mean
RMSE \NoDummyBaselineRMSELoss{} points higher
($p=\NoDummyBaselinePValue$), while Dummy RAPM has mean RMSE
\DummyBaselineRMSEGain{} points lower ($p=\DummyBaselinePValue$). The
improvement therefore appears to come from restoring information about lineup
composition rather than from excluding low-minute player coefficients.

\subsection{Dummy coefficients}

\begin{table}[htbp]
\centering
\small
\caption{Selected outer-training dummy-category exposure across seasons}
\label{tab:exposure}
\input{category_exposure_table.tex}
\end{table}

Table~\ref{tab:exposure} reports how often each home and away count occurs in
the selected outer-training designs. Exposure is the sum of
$q_i=(P_{H,i}+P_{A,i})/2$ over contributing stints. The season lists distinguish
structural zeros caused by an absent category from coefficients merely shrunk
close to zero.

\begin{table}[htbp]
\centering
\caption{Selected dummy coefficients across seasons}
\label{tab:coefficients}
\input{dummy_coefficient_table.tex}
\end{table}

The dominant coefficient pattern is straightforward. Conditional on the
retained-player indicators, a lineup with one excluded home player is
associated with
\HomeOneCoefficient{} home-margin points per 100 possessions. One excluded
away player is associated with \AwayOneCoefficient{} home-margin points.
Both signs are consistent with the same basketball interpretation: the side
using a low-minute player tends to be weaker. The home coefficient is negative in
\HomeOneNegativeSeasons{} seasons, and the away coefficient is positive in
\AwayOnePositiveSeasons{}.

\begin{figure}[htbp]
\centering
\includegraphics[width=0.68\linewidth]{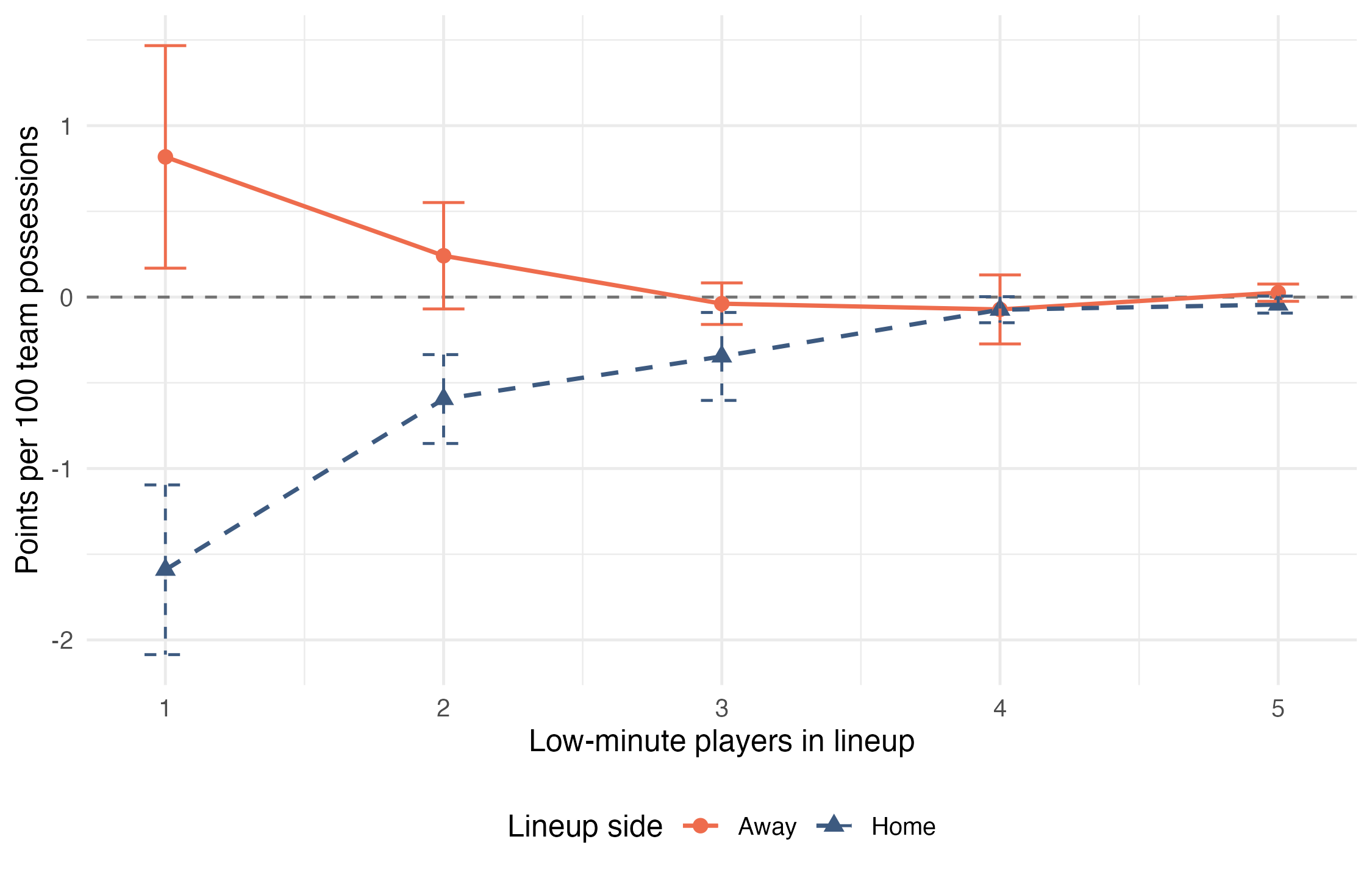}
\caption{Mean selected home and away dummy coefficients with 95\% cross-season
$t$ intervals. Away coefficients retain the home-margin
orientation.}
\label{fig:coefficients}
\end{figure}

These are lineup associations, not ratings of individual excluded players.
Score state, injuries, matchups, and schedule conditions all affect when
low-minute lineups appear. Counts of four and five occur in only a few seasons;
when a category is absent, its coefficient is structurally zero. The clearest
recurring pattern is concentrated in the one- and two-player categories. The
intervals in Figure~\ref{fig:coefficients} summarize variation across seasons.

\section{Discussion}

Why does this modest modification improve prediction so consistently? Most of
the predictive signal is already carried by the retained-player indicators.
Dropping low-minute columns discards only a limited amount of lineup
information, so a large reduction in RMSE should not be expected. Nevertheless,
the omitted information is predictive. Ten count indicators recover enough of
it to improve prediction in \RMSEWins{} of \NumberSeasons{} seasons.

The comparison with the unfiltered benchmark sharpens the point. Filtering
players while omitting their lineup counts makes the model slightly worse.
Recording those counts makes it slightly better. RAPM does not need a
separate, noisy coefficient for every low-minute player, but it benefits from
knowing how many such players are on the floor.

The selected ratio, $\rho=\SelectedPenaltyRatio$, places
\SelectedPenaltyRatio{} times as much penalty weight on the squared count
coefficients as on the squared player coefficients. Because predictors are not
standardized and occur at different frequencies, $\rho$ is not an effective
shrinkage factor and should not be read as the literal relative contraction of
the fitted coefficients. Nearby ratios perform nearly identically, so the
precise value of \SelectedPenaltyRatio{} is less important than the broader
result: the data favor a distinct, relatively strong penalty for the count
coefficients rather than their exclusion.

The remaining weaknesses come mostly from the data. Lineups are reconstructed
from play-by-play and starter records rather than an official shift chart.
Starter-data exclusions are concentrated in 2019, although the result survives
without that season. Categories with four or five low-minute players are rare,
so they contribute little to the substantive pattern.

\section{Conclusion}

When low-minute players are filtered out, they disappear from the design
matrix but not from the game. Five home and five away count indicators preserve
that lineup information. Across \NumberSeasons{} seasons, this representation
improves game-margin RMSE in \RMSEWins{} seasons. RAPM can therefore pool
low-minute players without discarding information about their presence on the
floor.

\section*{Reproducibility}

Code and instructions for reproducing the data construction, model fitting,
tables, figures, and manuscript are available at
\url{https://github.com/whartonsabi/dummy-rapm}.

\section*{Acknowledgments}

We thank the participants in the Wharton Sports Research Seminar for helpful
discussion and feedback.

{\footnotesize
\setlength{\bibsep}{4pt}
\bibliographystyle{apalike}
\bibliography{references}
}

\end{document}

%% file: results.tex
\newcommand{\SelectedThreshold}{10}
\newcommand{\SelectedPenaltyRatio}{2.2}
\newcommand{\CheckedPenaltyMinimum}{1}
\newcommand{\CheckedPenaltyMaximum}{2.5}

\newcommand{\NumberSeasons}{16}
\newcommand{\NumberSeasonsExcludingTwentyNineteen}{15}
\newcommand{\MeanDummyRMSE}{12.856}
\newcommand{\MeanNoDummyRMSE}{12.897}
\newcommand{\MeanRMSEGain}{0.042}
\newcommand{\RelativeRMSEGain}{0.30\%}
\newcommand{\RMSEGainCILow}{0.013}
\newcommand{\RMSEGainCIHigh}{0.070}
\newcommand{\RMSEWins}{13}
\newcommand{\NoDummyRMSEWins}{3}
\newcommand{\PairedTPValue}{0.0068}
\newcommand{\PairedTStatistic}{3.138}
\newcommand{\PairedTDegreesFreedom}{15}
\newcommand{\WilcoxonPValue}{0.0077}
\newcommand{\WilcoxonStatistic}{120.0}
\newcommand{\SignTestPValue}{0.0213}
\newcommand{\SignTestSuccesses}{13}
\newcommand{\MeanDummyRSquared}{0.178}
\newcommand{\MeanNoDummyRSquared}{0.173}

\newcommand{\HomeOneCoefficient}{-1.59}
\newcommand{\AwayOneCoefficient}{0.82}
\newcommand{\HomeOneNegativeSeasons}{16}
\newcommand{\AwayOnePositiveSeasons}{13}
\newcommand{\StarterExcludedGames}{33}
\newcommand{\StarterExcludedTwentyNineteenGames}{19}
\newcommand{\TotalModeledGames}{19,589}
\newcommand{\TotalModeledStints}{496,575}
\newcommand{\PreTwentyTwentyWins}{9}
\newcommand{\PreTwentyTwentySeasons}{12}
\newcommand{\PostCovidWins}{4}
\newcommand{\PostCovidSeasons}{4}
\newcommand{\ExcludingTwentyNineteenRMSEGain}{0.035}
\newcommand{\ExcludingTwentyNineteenPValue}{0.0132}
\newcommand{\NoDummyBaselineRMSELoss}{0.015}
\newcommand{\NoDummyBaselinePValue}{0.1129}
\newcommand{\DummyBaselineRMSEGain}{0.026}
\newcommand{\DummyBaselinePValue}{0.0626}

%% file: sample_summary_table.tex
\resizebox{\linewidth}{!}{%
\begin{tabular}{rrrrrrrrrr}
\toprule
Season & All & Oct--Dec & Jan--Feb & Oct--Feb & Mar--Apr & Out & Starter excl. & Stints & Players \\
\midrule
2007 & 1214 & 443 & 404 & 847 & 362 & 0 & 5 & 28,257 & 454 \\
2008 & 1217 & 448 & 411 & 859 & 358 & 0 & 0 & 27,243 & 451 \\
2009 & 1228 & 471 & 409 & 880 & 348 & 0 & 0 & 26,348 & 445 \\
2010 & 1225 & 470 & 409 & 879 & 346 & 0 & 0 & 26,835 & 441 \\
2011 & 1226 & 479 & 408 & 887 & 339 & 0 & 0 & 27,583 & 452 \\
2013 & 1226 & 453 & 402 & 855 & 366 & 0 & 5 & 23,607 & 469 \\
2014 & 1225 & 464 & 403 & 867 & 357 & 0 & 1 & 23,700 & 482 \\
2015 & 1226 & 477 & 393 & 870 & 356 & 0 & 0 & 25,134 & 492 \\
2016 & 1227 & 489 & 399 & 888 & 339 & 0 & 0 & 25,155 & 477 \\
2017 & 1230 & 506 & 387 & 893 & 336 & 0 & 1 & 25,032 & 484 \\
2018 & 1226 & 540 & 375 & 915 & 309 & 0 & 2 & 34,393 & 539 \\
2019 & 1230 & 532 & 376 & 908 & 303 & 0 & 19 & 34,502 & 537 \\
2022 & 1230 & 527 & 394 & 921 & 309 & 0 & 0 & 42,133 & 605 \\
2023 & 1230 & 546 & 385 & 931 & 299 & 0 & 0 & 42,531 & 539 \\
2024 & 1231 & 481 & 405 & 886 & 345 & 0 & 0 & 41,552 & 572 \\
2025 & 1231 & 485 & 403 & 888 & 343 & 0 & 0 & 42,570 & 569 \\
\bottomrule
\end{tabular}%
}

%% file: category_exposure_table.tex
\begin{tabular}{lrrrrp{5.0cm}}
\toprule
Side & Count & Stints & Exposure & Seasons & Seasons present \\
\midrule
Home & 0 & 341,583 & 1,256,501.5 & 16 & 2007--2011, 2013--2019, 2022--2025 \\
Home & 1 & 20,320 & 61,577.5 & 16 & 2007--2011, 2013--2019, 2022--2025 \\
Home & 2 & 2,476 & 7,957.5 & 16 & 2007--2011, 2013--2019, 2022--2025 \\
Home & 3 & 676 & 2,759.5 & 12 & 2007, 2013--2019, 2022--2025 \\
Home & 4 & 255 & 1,081.5 & 6 & 2018, 2019, 2022--2025 \\
Home & 5 & 45 & 202.0 & 4 & 2022--2025 \\
Away & 0 & 339,479 & 1,251,045.0 & 16 & 2007--2011, 2013--2019, 2022--2025 \\
Away & 1 & 21,903 & 65,940.5 & 16 & 2007--2011, 2013--2019, 2022--2025 \\
Away & 2 & 2,899 & 9,120.5 & 16 & 2007--2011, 2013--2019, 2022--2025 \\
Away & 3 & 763 & 2,702.0 & 13 & 2007, 2010, 2013--2019, 2022--2025 \\
Away & 4 & 266 & 1,090.5 & 7 & 2016, 2018, 2019, 2022--2025 \\
Away & 5 & 45 & 181.0 & 4 & 2022--2025 \\
\bottomrule
\end{tabular}

%% file: dummy_coefficient_table.tex
\begin{tabular}{lrrr}
\toprule
Lineup side & Low-minute players & Mean coefficient & Directional seasons \\
\midrule
Home & 1 & -1.59 & 16 negative \\
Home & 2 & -0.59 & 16 negative \\
Home & 3 & -0.35 & 12 negative, 4 zero \\
Home & 4 & -0.07 & 5 negative, 1 positive, 10 zero \\
Home & 5 & -0.04 & 4 negative, 12 zero \\
Away & 1 & 0.82 & 3 negative, 13 positive \\
Away & 2 & 0.24 & 7 negative, 9 positive \\
Away & 3 & -0.04 & 11 negative, 2 positive, 3 zero \\
Away & 4 & -0.07 & 4 negative, 3 positive, 9 zero \\
Away & 5 & 0.03 & 1 negative, 3 positive, 12 zero \\
\bottomrule
\end{tabular}